# Title: Observation of Fractionally Quantized Anomalous Hall Effect

**Authors:** Heonjoon Park[1†], Jiaqi Cai[1†], Eric Anderson[1†], Yinong Zhang[1], Jiayi Zhu[1], Xiaoyu Liu[2], Chong Wang[2], William Holtzmann[1], Chaowei Hu[1], Zhaoyu Liu[1], Takashi Taniguchi[5], Kenji Watanabe[6], Jiun-haw Chu[1], Ting Cao[2], Liang Fu[7], Wang Yao[3,4], Cui-Zu Chang[8], David Cobden[1], Di Xiao[2,1], Xiaodong Xu[1,2*]

[1]Department of Physics, University of Washington, Seattle, Washington 98195, USA
[2]Department of Materials Science and Engineering, University of Washington, Seattle, Washington 98195, USA
[3]Department of Physics, University of Hong Kong, Hong Kong, China
[4]HKU-UCAS Joint Institute of Theoretical and Computational Physics at Hong Kong, China
[5]Research Center for Materials Nanoarchitectonics, National Institute for Materials Science, 1-1 Namiki, Tsukuba 305-0044, Japan
[6]Research Center for Electronic and Optical Materials, National Institute for Materials Science, 1-1 Namiki, Tsukuba 305-0044, Japan
[7]Department of Physics, Massachusetts Institute of Technology, Cambridge, Massachusetts 02139, USA
[8]Department of Physics, The Pennsylvania State University, University Park, Pennsylvania, 16802, USA
[†] These authors contributed equally to the work.
*Corresponding author's email: xuxd@uw.edu

**Abstract:** The integer quantum anomalous Hall (QAH) effect is a lattice analog of the quantum Hall effect at zero magnetic field. This striking transport phenomenon occurs in electronic systems with topologically nontrivial bands and spontaneous time-reversal symmetry breaking. Discovery of its putative fractional counterpart in the presence of strong electron correlations, i.e., the fractional quantum anomalous Hall (FQAH) effect, would open a new chapter in condensed matter physics. Here, we report the direct observation of both integer and fractional QAH effects in electrical measurements on twisted bilayer $MoTe_2$. At zero magnetic field, near filling factor $\nu = -1$ (one hole per moiré unit cell) we see an extended integer QAH plateau in the Hall resistance $R_{xy}$ that is quantized to $h/e^2 \pm 0.1\%$ while the longitudinal resistance $R_{xx}$ vanishes. Remarkably, at $\nu = -\frac{2}{3}$ and $-\frac{3}{5}$ we see plateau features in $R_{xy}$ at $\frac{3}{2}h/e^2 \pm 1\%$ and $\frac{5}{3}h/e^2 \pm 3\%$, respectively, while $R_{xx}$ remains small. All these features shift linearly in an applied magnetic field with slopes matching the corresponding Chern numbers $-1$, $-\frac{2}{3}$, and $-\frac{3}{5}$, precisely as expected for integer and fractional QAH states. In addition, at zero magnetic field, $R_{xy}$ is approximately $2h/e^2$ near half filling ($\nu = -\frac{1}{2}$) and varies linearly as $\nu$ is tuned. This behavior resembles that of the composite Fermi liquid in the half-filled lowest Landau level of a two-dimensional electron gas at high magnetic field. Direct observation of the FQAH and associated effects paves the way for researching charge fractionalization and anyonic statistics at zero magnetic field.

**Main text:**

Topological materials are quantum phases of matter characterized by a bulk electronic band structure that exhibits nontrivial topological properties. The classic example is the integer quantum Hall effect. When a non-interacting two-dimensional electron gas is subjected to a large magnetic field, Landau levels form and the topology of the system changes[1]. As pointed out by Thouless *et al.*,[2] the topology is characterized by an invariant, known as the Chern number *C*, which determines the number of topologically protected chiral channels on the edge of the sample. When strong electron-electron interactions are present, fractional quantum Hall states can also emerge. These fractional states are fascinating in part because they can host exotic excitations which obey fractional statistics[3-11].

Both integer and fractional quantum Hall states require time-reversal symmetry breaking provided by an external magnetic field. However, Haldane predicted the existence of integer quantum Hall-like states at zero magnetic field[12], referred to as quantum anomalous Hall (QAH) states[13]. Unlike the conventional quantum Hall effect, these states do not rely on Landau-level formation. The integer QAH effect was first experimentally realized in magnetically doped topological insulator thin films[14]. More recently, it has also been observed in topological insulators with intrinsic magnetism[15] and in synthetic moiré quantum materials[16-20]. The hypothesized fractional QAH (FQAH) effects, which may occur in topological bands with strong electron correlations[21-26], had not been observed until the recent progress in twisted $MoTe_2$.[19,27,28]

Rhombohedral (R)-stacked twisted $MoTe_2$ bilayer, a 2D semiconductor hosting a honeycomb moiré superlattice, has emerged as a promising platform for investigating both QAH and FQAH effects[19,27,29-36]. Gate tunable correlated magnetic insulating states at both integer and fractional filling (*v*) of the moiré unit cell have been observed for twist angles in the range of 3º to 4º (Ref.[19,27]). The ferromagnetic order provides the spontaneous time-reversal symmetry breaking necessary for the realization of QAH and FQAH effects. Utilizing a trion sensing technique[19], a "fan diagram" was obtained by tracking the three ferromagnetic insulating states (*v* = -1, -2/3, and -3/5) in the hole density-magnetic field plane. Remarkably, these three states were found to evolve with slopes corresponding to Chern numbers *C* = -1, -2/3, and -3/5 respectively, consistent with the characteristics of integer and fractional QAH states according to the Streda formula. The *C* = -1 and -2/3 behavior have also since been confirmed through exciton sensing measurements[28], indicating the robustness of both the phenomena and the material platform. Nevertheless, the fundamental defining feature of a QAH state is a quantized Hall resistance indicating dissipationless chiral edge modes and insulating bulk at zero magnetic field.

Here, we report transport measurements which provide direct evidence of both the integer and fractional QAH states in this system. From these measurements, we establish a topological phase diagram as a function of doping and perpendicular electric field. Historically, the main difficulty in twisted 2D semiconductor transport studies[37,38] has been achieving low contact resistance at zero perpendicular electric field. To overcome this, we utilize the scheme shown in Fig. 1a, incorporating additional gates to lower the contact resistance by heavily hole-doping the contact region to a carrier density of about $10^{13}$ cm$^{-2}$ (See Methods). Data presented in the main text are from two devices denoted by D(θ), with the twist angles θ = 3.7º and 3.9º, respectively (Extended Data Fig. 1). Data from an additional device D(3.52º) is presented in Extended Data Fig. 2.

**Identification of QAH states**

Figures 1b and c show maps of longitudinal ($R_{xx}$) and Hall ($R_{xy}$) resistance measurements on device D(3.7°) as a function of doping $n$ and perpendicular electric field $D/\varepsilon_0$ at the temperature of $T$ = 100 mK (Methods). $R_{xx}$ is symmetrized and $R_{xy}$ is antisymmetrized with respect to a small applied out-of-plane magnetic field ($\mu_0 H$) of ±200 mT. The field serves to stabilize the polarization of the ferromagnetic states when sweeping the gates. Insulating behavior, with $R$ > 1 MΩ, is shown in black. The filling factor $\nu$ is indicated on the top axis. Multiple intriguing gate-tunable features can be seen in these maps which call for theoretical investigation. Here, we focus on the regions centered around $\nu$ = -1, -2/3, and -1/2 and $D/\varepsilon_0$ = 0. In these regions of phase space, there are areas where $R_{xx}$ is small and $|R_{xy}|$ is very large, implying the emergence of topologically non-trivial states.

The nontrivial topological nature of these states is immediately evident in the magnetic field dependence of $R_{xy}$. As shown in Fig. 1d, the states at $\nu$ = -1, -2/3, and -3/5 shift linearly with $|\mu_0 H|$ down to zero magnetic field. The slopes match the overlaid dashed lines, which are the expected dispersions of states with nonzero Chern numbers $C$ according the Streda formula. The broad band of uniform $R_{xy}$ centered on $\nu$ = -1 shifts with a slope given by $|C|$ = 1, consistent with a QAH state. The narrower maxima in $R_{xy}$ located at $\nu$ = -2/3 and $\nu$ = -3/5 when $|\mu_0 H|$ = 0 shift with slopes given by $|C|$ = 2/3 and 3/5, respectively, indicative of FQAH states. The slopes change sign when $\mu_0 H$ is reversed because the system magnetization flips, thus changing the sign of the Chern numbers. The observed states are consistent with the prior results obtained via trion sensing[19].

As elucidated in prior reports[19,27], R-stacked MoTe$_2$ contains two degenerate moiré orbitals. These orbitals are localized in opposite layers and sit on the two triangular sublattices of a honeycomb lattice, realizing the Kane-Mele model (Fig. 1e&f).[31,32] Figure 1f shows the calculated spin/valley-resolved moiré valence bands in the presence of spontaneous time reversal symmetry breaking (see Methods). The associated Chern numbers of the bands have opposite signs for the opposite valleys. Depending on the filling of the top topological moiré flat band, both integer and fractional QAH states can form. Below, we present a detailed investigation of these states.

**Integer QAH state**

Figure 2a shows the magnetic field dependence of $R_{xy}$ and $R_{xx}$ of device D(3.9°) at $\nu$ = -1 for selected values of $D/\varepsilon_0$ at 100 mK. At $D/\varepsilon_0$=0, $R_{xy}$ exhibits a hysteresis loop with quantized values ±$h/e^2$ (0.9998±0.0136× $h/e^2$) at $\mu_0 H$ = 0 and coercive field $\mu_0 H_c$ ~ 84 mT. Correspondingly, $R_{xx}$ nearly vanishes at $\mu_0 H$ = 0 and shows peaks of height ≈ 0.3 $h/e^2$ when $R_{xy}$ jumps between the quantized values. This is precisely the hallmark of the QAH effect. The QAH state remains robust as electric field increases, and starts to diminish as $D/\varepsilon_0$ rises above a critical value. Note that the exact value of critical electric field depends on the twist angle (e.g. see Extended Data Fig. 2 for D(3.52°)).

To demonstrate this electric field effect, Fig. 2b shows the antisymmetrized $R_{xy}$ and symmetrized $R_{xx}$ as a continuous function of $D/\varepsilon_0$ at $|\mu_0 H|$ = 200 mT. At the critical electric field $D_c/\varepsilon_0$ ≈ 150 mV/nm, $R_{xy}$ begins to fall and $R_{xx}$ starts to rise (see also Extended Data 3). By $D/\varepsilon_0$ ≈ 180 mV/nm, $R_{xy}$ has fallen to nearly zero and $R_{xx}$ exceeds 100 kΩ. This behavior implies an electric field-induced phase transition from a QAH state to a trivial insulating state. This phase transition appears to be continuous, judging by the absence of hysteresis seen in $R_{xx}$ when the electric field is cycled through the critical value. Similar phenomena were observed in device D(3.52°) (Extended Data Fig. 2), where combined reflective magnetic circular dichroism and transport measurements show that this topologically trivial insulating state remains ferromagnetic. This finding is corroborated by Hartree-Fock calculations performed as a function of electric field at $v = -1$ (Extended Data Fig. 2; see Methods), as well as other very recent theoretical work.[39,40]

We can learn more about the phase transition between the QAH and ferromagnetic insulator states by studying the temperature dependence of the resistance. Figure 2c shows the temperature dependence of $\Delta R_{xy}$, the difference between $R_{xy}$ values when magnetic field is swept down and up (see Extended Data Fig. 3). Remarkably, $\Delta R_{xy}$ is nearly quantized even at $T = 8$ K, near the Curie temperature of ~12 K. The temperature and electric field dependence of $R_{xx}$ at $\mu_0 H = 100$ mT is shown in Fig. 2d, where the color scale is chosen to emphasize the transition occurring near 150 mV/nm.

We estimate an energy gap $\Delta$ from the data at given $v$ and $D/\varepsilon_0$ using Arrhenius plots[14,15,17,41] (see also Extended Data Fig. 3). For $D/\varepsilon_0 < D_c/\varepsilon_0$, in the QAH state, we anticipate that for small values of $R_{xx}$ the resistance is thermally activated, and thus fit $R_{xx}$ to $R_0 e^{-\Delta/2k_B T}$ for low T ($k_B$ is the Boltzmann constant). For $D/\varepsilon_0 > D_c/\varepsilon_0$, in the insulating state, we expect thermally activated longitudinal conductivity, and hence fit $R_{xx}$ to $R_0 e^{+\Delta/2k_B T}$ instead. For $D/\varepsilon_0$ near $D_c/\varepsilon_0$, $R_{xx}$ is almost independent of $T$ and of order $h/e^2$.[42] This behavior resembles that observed at quantum Hall liquid to Hall insulator[43] and QAH to Anderson insulator[44] phase transitions.

Figure 2e plots the extracted values of $\Delta$ as a function of $D/\varepsilon_0$. We find that in the QAH state, $\Delta$ initially decreases gradually as $D/\varepsilon_0$ is increased from zero before shrinking rapidly as $D/\varepsilon_0$ approaches $D_c/\varepsilon_0$. Near $D/\varepsilon_0 = D_c/\varepsilon_0$, $\Delta$ is small, but its value cannot be determined precisely as $R_{xx}$ is nearly independent of temperature. As $D/\varepsilon_0$ is increased beyond $D_c/\varepsilon_0$, $\Delta$ increases again as the system enters the topologically trivial insulating phase. This change of $\Delta$ with increasing $D/\varepsilon_0$ strongly suggests a continuous closing and reopening of an energy gap, as expected for a continuous topological quantum phase transition. The results of a similar analysis of the Hall resistance are shown in Extended Data. Fig. 3. We note that the integer QAH states are robust over a large range of twist angles (from around 3.5 to 3.9°, see Extended Data Fig. 1). Therefore, we expect routine studies of robust QAH physics in 2D materials to become feasible using twisted MoTe$_2$.

**FQAH states**

We now turn to the FQAH states. All data shown are from device D(3.7°). Figure 3a shows the magnetic field dependence of $R_{xx}$ and $R_{xy}$ at $v = -2/3$, with $T = 500$ mK and $D/\varepsilon_0 = 0$. The measurement configuration is detailed in Extended Data Fig. 4. $R_{xy}$ exhibits a hysteresis loop, with coercive field $\mu_0 H_c \sim 20$ mT, between levels at $\pm 3h/2e^2$ within 1% accuracy. Meanwhile, $R_{xx}$ remains below 1 k$\Omega$ except for small bumps when $R_{xy}$ switches between the two quantized values. These results confirm that the system at $v = -2/3$ is in a FQAH state with Chern number $C = -2/3$. Figure 3b shows similar behavior at $v = -3/5$, where $R_{xy}$ exhibits a hysteresis loop between levels at $\pm 5h/3e^2$ within 3% accuracy and $R_{xx}$ again is small, confirming that this is a FQAH state with $C = -3/5$.

Like the $C = -1$ QAH state, the FQAH states are electrically tunable. For simplicity, we focus on investigation of the -2/3 FQAH state. Figure 3c shows the extracted antisymmetrized $R_{xy}$ and symmetrized $R_{xx}$ at $|\mu_0 H| = 50$ mT as a continuous function of $D/\varepsilon_0$ at $T = 100$ mK. $R_{xy}$ starts to decrease above the critical electric field $|D_c/\varepsilon_0| \sim 20$ mV/nm and eventually disappears into noise as $|D/\varepsilon_0|$ rises above 80 mV/nm. Correspondingly, $R_{xx}$ starts to rise from zero at ~50 mV/nm, reaches a peak in the range 100-200 k$\Omega$ roughly where $R_{xy}$ disappears, and then drops to 20-30 k$\Omega$ as the electric field further increases to about 150 mV/nm. This behavior implies an electric field induced phase transition from a FQAH state to a topologically trivial resistive state, and eventually to a metallic state. This phase transition can also be observed in Figs. 1b and c. Further theoretical and experimental efforts are needed to understand the properties of these electronic phases at large $D$.

To estimate the energy gap of the FQAH states, we investigate their temperature dependence. Figure 3d shows the magnetic field dependence of $R_{xy}$ and $R_{xx}$ for the -2/3 state at selected temperatures. Figure 3e is a color plot of $\Delta R_{xy}$ versus magnetic field and temperature. As $T$ is increased above 1 K, $R_{xy}$ starts to deviate from the quantized value at zero magnetic field, though it remains quantized under a small applied magnetic field (>20 mT) up to ~2 K. Above this temperature quantization is lost, until the hysteresis eventually vanishes for $T > 4$ K. Extended Data. Fig. 5 shows Arrhenius plots of $R_{xx}$. The energy gap of the -2/3 FQAH state is estimated by fitting the data to $R_0 e^{-\Delta/2k_B T}$, which yields 23(7) K. A similar analysis performed for the thermally activated ($3h/2e^2$-$R_{xy}$) is consistent with this value. Using the same procedure (see Extended Data Fig. 6), we find a smaller gap of $\Delta = 15(3)$ K for the -3/5 FQAH state. This is consistent with the conclusions of the optical measurements[19] and exact diagonalization calculations[29].

**Half-filled QAH state**

Finally, we examine the behavior near half filling in device D(3.7°). Similar results from D(3.9°) are included in Extended Data Fig. 7. Figures 4a and b show the magnetic field dependence of $R_{xx}$ and $R_{xy}$ at $D/\varepsilon_0 = 0$, T = 500 mK, and near $v = -1/2$. $R_{xy}$ exhibits hysteresis with coercive field $\mu_0 H_c$ ~16 mT and has a value of about $\pm 2h/e^2$ at zero magnetic field. Meanwhile, $R_{xx}$ is approximately 2 k$\Omega$, apart from a small increase when $R_{xy}$ flips. The $v$ dependence of $R_{xx}$ and $R_{xy}$ at 100 mK, symmetrized and antisymmetrized at $|\mu_0 H| = 50$ mT, is shown in Fig. 4c. We see that the quantized values close to $3h/2e^2$ and $5h/3e^2$ obtained at $v = -2/3$ and -3/5 are located at the tops of humps in $R_{xy}$ vs $v$. Near half filling, $R_{xy}$ becomes almost linear in $v$, and passes through a value close to $2h/e^2$ at $v = -1/2$. Across the entire range, $R_{xx}$ remains at just a few k$\Omega$. The linear dependence of $R_{xy}$ vs $v$ suggests that the state is compressible, explaining why no gapped feature at $v = -1/2$ was seen in optical (trion sensing) measurements[19,27]. This behavior of $R_{xx}$ and $R_{xy}$ bears a clear resemblance to that observed near half filling of the lowest Landau level in a two-dimensional electron gas at high magnetic field, which has been found to be a composite Fermi liquid (CFL) state[45-47]. Indeed, recent exact diagonalization calculations have predicted that a zero-field CFL state is the ground state at $v = -1/2$ in twisted MoTe$_2$[39,48].

We additionally performed temperature dependent measurements of $R_{xx}$ and $R_{xy}$. Extended Data Fig. 8 shows $\Delta R_{xy}$, the hysteretic component of $R_{xy}$, versus temperature. $\Delta R_{xy}$ vanishes for $T$ above ~4 K. Temperature dependent $R_{xx}$ and $R_{xy}$ near $v = -1/2$ are plotted in Fig. 4d. Here $R_{xx}$ and $R_{xy}$ are symmetrized and antisymmetrized at $|\mu_0 H| = 50$ mT. As temperature increases, the $R_{xy}$ plateau of about $2h/e^2$ gradually tapers off around 1.7 K and goes to zero above 4 K. Concurrently, $R_{xx}$ initially around 2 kΩ starts to increase around 2 K. This behavior may reflect a temperature-driven phase transition from a half-filled QAH state to a topologically trivial higher temperature state.

In summary, we provide electrical transport measurements directly confirming that R-stacked twisted MoTe$_2$ bilayer hosts both integer ($C$ = -1) and fractional ($C$ = -2/3 and -3/5) QAH states, the presence of which were first indicated by thermodynamic arguments applied to recent optical measurements[19]. In addition, a new half-filled QAH state was observed. This state was not seen in the optical measurements, likely due to its compressible nature. Future experiments are needed to establish whether the QAH state at half filling is truly a zero-field CFL – a non-Fermi liquid state resulting from the interplay between strong interactions and spontaneous time reversal symmetry breaking. For example, it has been suggested that the CFL state[48] could exhibit intrinsic commensurability oscillations of the resistance as $v$ varies near -1/2. This phenomenon, related to the commensurability oscillations in a Landau level system where an external periodic potential is imposed (Ref. [49-52]), would be unique to the zero-field CFL state in moiré superlattices, having a characteristic dependence on moiré wavelength, i.e., twist angle. Anticipating further improvements in crystal quality and contact technology, we expect the family of twisted MoTe$_2$ systems to be a powerful playground for exploring the interplay between correlations, topology, and magnetism, and to serve as a versatile platform for topological electronic and spintronic devices.

**Methods:**

**Sample fabrication.** hBN, 2H-MoTe$_2$, and graphite crystals were mechanically exfoliated onto p-doped Si/SiO$_2$ substrates to obtain nm thick flakes. An optical microscope was used to find hBN for top, bottom, and contact gate dielectrics, and thin graphite for gate electrodes. The cleanliness of the flakes was confirmed by atomic force microscopy (AFM) or contrast-enhanced optical microscopy. The bottom gate heterostructure was fabricated by a standard polycarbonate-based dry transfer process. First, an hBN flake was chosen to serve as a bottom gate dielectric and picked up, followed by a graphite bottom gate electrode, before putting down the structure on a silicon substrate. Standard e-beam lithography was used to define Hall bar contacts, and Pt (8 nm) electrodes were deposited by e-beam evaporation. The bottom gate structure was cleaned using contact-mode AFM, and then moved into a glovebox with O$_2$ and H$_2$O levels < 0.1 ppm. In the glovebox, MoTe$_2$ was exfoliated to find a monolayer flake which was cut using an AFM tip. A twisted MoTe$_2$ stack was formed by picking up an hBN flake, picking up part of the MoTe$_2$ monolayer, rotating the stage by a desired angle, and then picking up the remaining part of the MoTe$_2$ monolayer. The twisted MoTe$_2$ structure was put down onto the prepared bottom gate. The stamp polymer was washed off with molecular sieve-cleaned anhydrous chloroform and dichloromethane in a glovebox environment. After finishing the encapsulation and wash-off process, the twisted MoTe$_2$ device was removed from the glovebox and cleaned using contact mode AFM. Local contact gates, used to improve the contact resistance, were added using standard e-beam lithography and evaporation of Pt (8 nm). An additional step of lithography and evaporation of Cr/Au (5/70 nm) was performed to connect electrodes and conductive pads to allow for wire-bonding to the sample. After lithography, another round of AFM cleaning was performed on the entire structure before the subsequent transfer. The structure was finished by adding a graphite/hBN top gate. Polymer from this transfer was washed off with chloroform in ambient conditions.

**Electrical measurements.** Transport measurements were conducted in a Bluefors dilution refrigerator with a 9 T magnet and a base phonon temperature ≈ 20 mK and electron temperature ≈ 80 mK, as estimated from measurements of graphene devices with similar geometry. All AC signals were generated and detected by SR830 lock-in amplifiers. We found that a constant current measurement scheme significantly improves the stability of the magnetic state compared to a constant voltage scheme. An AC current bias of ≈ 0.2-0.5 nA was generated using a 100 MΩ resistor in series with an AC voltage source, and the actual current was monitored using a DL1211 current preamplifier. Four-terminal $R_{xx}$ and $R_{xy}$ signals were amplified using the differential mode of an SR560 voltage preamplifier with an input resistance (≈ 100 MΩ) much larger than the device contact resistance. Two-terminal conductance was measured by applying a 200-500 μV AC voltage bias to the source and monitoring the drain current with a DL1211. For temperature dependences above 4 K the current on a resistive heater mounted near the sample was gradually ramped up, with the temperature read out using a Cernox sensor close to the sample. A test with a carbon film sensor mounted on the chip carrier showed that the temperature difference between the sensor and sample stage temperature does not exceed 50 mK.

**Reflective magnetic circular dichroism (RMCD) measurements.** RMCD measurements were performed in a closed-loop magneto-optical exchange gas cryostat (attoDRY 2100) with an attocube *xyz* piezo stage, a 9T out-of-plane superconducting magnet, and a base temperature of 1.6 K. RMCD excitation was achieved by filtering a broadband supercontinuum source (NKT SuperK FIANIUM FIU-15) via dual-passing through a monochromator, selecting out an energy resonant with the trion feature (1115 nm) of hole-doped twisted MoTe$_2$, as reported previously[27]. RMCD signal is proportional to the difference in reflectance of right and left circularly polarized (RCP and LCP) light normalized by the total reflectance. The excitation was chopped at 1 kHz and the polarization was alternated between RCP and LCP with a photoelastic modulator at 50 kHz. The reflected light was detected by an InGaAs avalanche photodiode and read out by two lock-in amplifiers (SR830) at 50 kHz and 1 kHz. This scheme allows for continuous detection of both the reflectance difference between RCP and LCP and the total reflectance, thus allowing RMCD to be extracted.

**Estimation of filling factor based on doping density.**
The carrier density $n$ and electric field $D$ on the sample were derived from top (bottom) gate voltages $V_{tg}$ and $V_{bg}$ using $n = (V_{tg}C_{tg} + V_{bg}C_{bg})/e - n_{offset}$ and $D/\varepsilon_0 = (V_{tg}C_{tg} - V_{bg}C_{bg})/2\varepsilon_0 - D_{offset}$, where $e$ is the electron charge, $\varepsilon_0$ is the vacuum permittivity, and $C_{tg}$ and $C_{bg}$ are the top and bottom gate capacitances obtained from the gate thickness measured by AFM. $D_{offset} \approx 0$ was inferred from the dual gate resistance map. For D(3.52$^0$), the specific filling factor $v = -2$ can be identified from the position of a prominent peak in $R_{xx}$, and $v = -1$ from a salient feature in the RMCD. For D(3.7$^0$) and D(3.9$^0$), the sequence of fractional fillings with prominent features in the dual gate resistance map (Fig. 1b,c) can be used to identify the moiré filling factor. The gate capacitances and assignments of filling factor were independently verified by high magnetic field Landau fan fitting at finite electric field, and by the Streda slope of $v = -1$ and $-2/3$ (in D(3.7$^0$) and D(3.9$^0$)) at zero electric field.

**Ground state calculation**

We start from the continuum model for twisted MoTe$_2$ bilayer introduced in Ref.[31]. As shown in Ref.[29], at a twist angle of 3.5° in the absence of an electric field, the continuum model realizes a Kane-Mele model with the first two moiré valence bands having Chern number -1 and 1. This allows us to construct a tight-binding Hamiltonian on the honeycomb lattice. With the inclusion of electric field, the single-particle Hamiltonian reads

$$H_{single-particle} = \sum_{\sigma}\left(\sum_{\langle i,j \rangle} t_1 c_{i\sigma}^\dagger c_{j\sigma} + \sum_{\langle\langle i,j \rangle\rangle} t_2 e^{i\nu_{ij}\theta} c_{i\sigma}^\dagger c_{j\sigma} + D \sum_{i \in A} c_{i\sigma}^\dagger c_{i\sigma}\right).$$

Here, $\sigma$ refers to spin, $t_1$ is the nearest-neighbor hopping between A and B sublattices, $t_2$ is the next-nearest neighbor hopping within a single sublattice, and $D$ is the interlayer potential difference introduced by applied electric field. $\langle i,j \rangle$ and $\langle\langle i,j \rangle\rangle$ refers to nearest neighbor and next-nearest neighbor hopping, respectively. $\nu_{ij} = \pm 1$ depends on the hopping direction (Ref.[53]). We choose $t_2 = 0.2 t_1$ and $\theta = \pi/3$. The onsite Hubbard interaction is chosen as $U = 10 t_1$. We perform Hartree-Fock calculations on this Hamiltonian at the filling factor $\nu = -1$ under different electric fields.

**References:**

1  Klitzing, K. v., Dorda, G. & Pepper, M. New Method for High-Accuracy Determination of the Fine-Structure Constant Based on Quantized Hall Resistance. *Physical Review Letters* **45**, 494-497 (1980).
2  Thouless, D. J., Kohmoto, M., Nightingale, M. P. & den Nijs, M. Quantized Hall Conductance in a Two-Dimensional Periodic Potential. *Physical Review Letters* **49**, 405-408 (1982).
3  Tsui, D. C., Stormer, H. L. & Gossard, A. C. Two-dimensional magnetotransport in the extreme quantum limit. *Physical Review Letters* **48**, 1559 (1982).
4  Laughlin, R. B. Anomalous quantum Hall effect: an incompressible quantum fluid with fractionally charged excitations. *Physical Review Letters* **50**, 1395 (1983).
5  Halperin, B. I. Statistics of Quasiparticles and the Hierarchy of Fractional Quantized Hall States. *Physical Review Letters* **52**, 1583-1586 (1984).
6  Arovas, D., Schrieffer, J. R. & Wilczek, F. Fractional Statistics and the Quantum Hall Effect. *Physical Review Letters* **53**, 722-723 (1984).
7  Zhang, S. C., Hansson, T. H. & Kivelson, S. Effective-field-theory model for the fractional quantum Hall effect. *Physical review letters* **62**, 82 (1989).
8  Moore, G. & Read, N. Nonabelions in the fractional quantum Hall effect. *Nucl. Phys. B* **360**, 362-396 (1991).
9  Wen, X.-G. Non-Abelian statistics in the fractional quantum Hall states. *Physical review letters* **66**, 802 (1991).

10	Nakamura, J., Liang, S., Gardner, G. C. & Manfra, M. J. Direct observation of anyonic braiding statistics. *Nature Physics* **16**, 931-936 (2020).
11	Bartolomei, H. *et al.* Fractional statistics in anyon collisions. *Science* **368**, 173-177 (2020).
12	Haldane, F. D. M. Model for a quantum Hall effect without Landau levels: Condensed-matter realization of the "parity anomaly". *Physical Review Letters* **61**, 2015 (1988).
13	Chang, C.-Z., Liu, C.-X. & MacDonald, A. H. Colloquium: Quantum anomalous hall effect. *Reviews of Modern Physics* **95**, 011002 (2023).
14	Chang, C.-Z. *et al.* Experimental Observation of the Quantum Anomalous Hall Effect in a Magnetic Topological Insulator. *Science* **340**, 167-170 (2013).
15	Deng, Y. *et al.* Quantum anomalous Hall effect in intrinsic magnetic topological insulator $MnBi_2Te_4$. *Science* **367**, 895-900 (2020).
16	Sharpe, A. L. *et al.* Emergent ferromagnetism near three-quarters filling in twisted bilayer graphene. *Science* **365**, 605-608 (2019).
17	Serlin, M. *et al.* Intrinsic quantized anomalous Hall effect in a moiré heterostructure. *Science* **367**, 900-903 (2020).
18	Li, T. *et al.* Quantum anomalous Hall effect from intertwined moiré bands. *Nature* **600**, 641-646 (2021).
19	Cai, J. *et al.* Signatures of fractional quantum anomalous hall states in twisted mote2 bilayer. *Nature* https://www.nature.com/articles/s41586-023-06289-w (2023).
20	Foutty, B. A. *et al.* Mapping twist-tuned multi-band topology in bilayer $WSe_2$. *arXiv preprint arXiv:2304.09808* (2023).
21	Sheng, D., Gu, Z.-C., Sun, K. & Sheng, L. Fractional quantum Hall effect in the absence of Landau levels. *Nature Communications* **2**, 389 (2011).
22	Neupert, T., Santos, L., Chamon, C. & Mudry, C. Fractional quantum Hall states at zero magnetic field. *Physical Review Letters* **106**, 236804 (2011).
23	Tang, E., Mei, J.-W. & Wen, X.-G. High-temperature fractional quantum Hall states. *Physical Review Letters* **106**, 236802 (2011).
24	Regnault, N. & Bernevig, B. A. Fractional chern insulator. *Physical Review X* **1**, 021014 (2011).
25	Xie, Y. *et al.* Fractional Chern insulators in magic-angle twisted bilayer graphene. *Nature* **600**, 439-443 (2021).
26	Spanton, E. M. *et al.* Observation of fractional Chern insulators in a van der Waals heterostructure. *Science* **360**, 62-66 (2018).
27	Eric Anderson, F.-R. F., Jiaqi Cai, William Holtzmann, Takashi Taniguchi, Kenji Watanabe, Di Xiao, Wang Yao, and Xiaodong Xu. Programming Correlated Magnetic States via Gate Controlled Moiré Geometry. *Science* https://www.science.org/doi/full/10.1126/science.adg4268 (2023).
28	Zeng, Y. *et al.* Integer and fractional Chern insulators in twisted bilayer MoTe2. *arXiv preprint arXiv:2305.00973* (2023).
29	Wang, C. *et al.* Fractional Chern Insulator in Twisted Bilayer $MoTe_2$. *arXiv preprint arXiv:2304.11864* (2023).
30	Li, H., Kumar, U., Sun, K. & Lin, S.-Z. Spontaneous fractional Chern insulators in transition metal dichalcogenide moiré superlattices. *Physical Review Research* **3**, L032070 (2021).

31  Wu, F., Lovorn, T., Tutuc, E., Martin, I. & MacDonald, A. Topological insulators in twisted transition metal dichalcogenide homobilayers. *Physical Review Letters* **122**, 086402 (2019).
32  Yu, H., Chen, M. & Yao, W. Giant magnetic field from moiré induced Berry phase in homobilayer semiconductors. *National Science Review* **7**, 12-20 (2020).
33  Zhang, Y.-H., Mao, D., Cao, Y., Jarillo-Herrero, P. & Senthil, T. Nearly flat Chern bands in moiré superlattices. *Physical Review B* **99**, 075127 (2019).
34  Devakul, T., Crépel, V., Zhang, Y. & Fu, L. Magic in twisted transition metal dichalcogenide bilayers. *Nature Communications* **12**, 6730 (2021).
35  Reddy, A. P., Alsallom, F. F., Zhang, Y., Devakul, T. & Fu, L. Fractional quantum anomalous Hall states in twisted bilayer $MoTe_2$ and $WSe_2$. *arXiv preprint arXiv:2304.12261* (2023).
36  Crépel, V. & Fu, L. Anomalous Hall metal and fractional Chern insulator in twisted transition metal dichalcogenides. *arXiv preprint arXiv:2207.08895* (2022).
37  Ghiotto, A. *et al.* Quantum criticality in twisted transition metal dichalcogenides. *Nature* **597**, 345-349 (2021).
38  Wang, L. *et al.* Correlated electronic phases in twisted bilayer transition metal dichalcogenides. *Nature materials* **19**, 861-866 (2020).
39  Dong, J. a. W., Jie and Ledwith, Patrick J. and Vishwanath, Ashvin and Parker, Daniel E. Composite Fermi Liquid at Zero Magnetic Field in Twisted $MoTe_2$. *arXiv preprint arXiv:2306.01719* (2023).
40  Fu, T. W. a. T. D. a. M. P. Z. a. L. Topological magnets and magnons in twisted bilayer $MoTe_2$ and $WSe_2$. *arXiv preprint arXiv:2306.02501* (2023).
41  Wei, H. P., Chang, A. M., Tsui, D. C. & Razeghi, M. Temperature dependence of the quantized Hall effect. *Physical Review B* **32**, 7016-7019 (1985).
42  Kivelson, S., Lee, D.-H. & Zhang, S.-C. Global phase diagram in the quantum Hall effect. *Physical Review B* **46**, 2223 (1992).
43  Shahar, D., Tsui, D. C., Shayegan, M., Bhatt, R. N. & Cunningham, J. E. Universal Conductivity at the Quantum Hall Liquid to Insulator Transition. *Physical Review Letters* **74**, 4511-4514 (1995).
44  Chang, C.-Z. *et al.* Observation of the Quantum Anomalous Hall Insulator to Anderson Insulator Quantum Phase Transition and its Scaling Behavior. *Physical Review Letters* **117**, 126802 (2016).
45  Willett, R. *et al.* Observation of an even-denominator quantum number in the fractional quantum Hall effect. *Physical review letters* **59**, 1776 (1987).
46  Halperin, B. I., Lee, P. A. & Read, N. Theory of the half-filled Landau level. *Physical Review B* **47**, 7312 (1993).
47  Willett, R. L., Ruel, R. R., West, K. W. & Pfeiffer, L. N. Experimental demonstration of a Fermi surface at one-half filling of the lowest Landau level. *Physical Review Letters* **71**, 3846-3849 (1993).
48  Goldman, H., Reddy, A. P., Paul, N. & Fu, L. Zero-field composite Fermi liquid in twisted semiconductor bilayers. *arXiv preprint arXiv:2306.02513* (2023).
49  Weiss, D., Klitzing, K., Ploog, K. & Weimann, G. Magnetoresistance oscillations in a two-dimensional electron gas induced by a submicrometer periodic potential. *Europhysics Letters* **8**, 179 (1989).

50  Gerhardts, R., Weiss, D. & Klitzing, K. v. Novel magnetoresistance oscillations in a periodically modulated two-dimensional electron gas. *Physical review letters* **62**, 1173 (1989).
51  Winkler, R., Kotthaus, J. & Ploog, K. Landau band conductivity in a two-dimensional electron system modulated by an artificial one-dimensional superlattice potential. *Physical review letters* **62**, 1177 (1989).
52  Kang, W., Stormer, H. L., Pfeiffer, L. N., Baldwin, K. W. & West, K. W. How real are composite fermions? *Physical Review Letters* **71**, 3850-3853 (1993).
53  Kane, C. L. & Mele, E. J. $Z_2$ Topological Order and the Quantum Spin Hall Effect. *Physical Review Letters* **95**, 146802 (2005).

**Acknowledgements:** Measurements of the fractional QAH states are supported by DoE BES under award DE-SC0018171. Measurements of the integer QAH state are supported by AFOSR FA9550-21-1-0177. Device fabrication and electrical transport measurements are partially supported by the Center on Programmable Quantum Materials, an Energy Frontier Research Center funded by DOE BES under award DE-SC0019443. The understanding of magnetism and the topological phase diagram is partially supported by AFOSR Multidisciplinary University Research Initiative (MURI) program, grant no. FA9550- 19-1-0390. The authors also acknowledge the use of the facilities and instrumentation supported by NSF MRSEC DMR-1719797. EA acknowledges support by the National Science Foundation Graduate Research Fellowship Program under Grant No. DGE-2140004. WY acknowledges support from the Research Grants Council of Hong Kong SAR (AoE/P-701/20, HKU SRFS2122-7S05) and Croucher Foundation. C. -Z. C. also acknowledges the support from Gordon and Betty Moore Foundation's EPiQS Initiative (GBMF9063). K.W. and T.T. acknowledge support from the JSPS KAKENHI (Grant Numbers 20H00354, 21H05233 and 23H02052) and World Premier International Research Center Initiative (WPI), MEXT, Japan. This research acknowledges usage of the millikelvin optoelectronic quantum material laboratory supported by the M. J. Murdock Charitable Trust. JHC and XX acknowledge support from the State of Washington funded Clean Energy Institute.

**Author contributions:** XX conceived and supervised the project. HP, JC, and EA fabricated and characterized the samples, assisted by YZ, JZ, WH, ZL, CH and JHC. HP and JC performed the transport measurements with inputs from CZC. EA performed the magnetic circular dichroism measurements. JC, DC, and XX provided dilution fridge measurement support and designed the contact schemes. HP, JC, EA, DC, CZC, TC, LF, WY, DX, and XX analyzed and interpreted the results. XL, CW, TC, and DX performed Hartree-Fock calculations. TT and KW synthesized the hBN crystals. HP, JC, EA, LF, CZC, DC, DX, and XX wrote the paper with input from all authors. All authors discussed the results.

**Competing Interests:** XX, EA, HP, and JC have applied for a patent partially based on this work. The other authors declare no competing interests.

**Data Availability:** The datasets generated during and/or analyzed during this study are available from the corresponding author upon reasonable request.

# Figures:

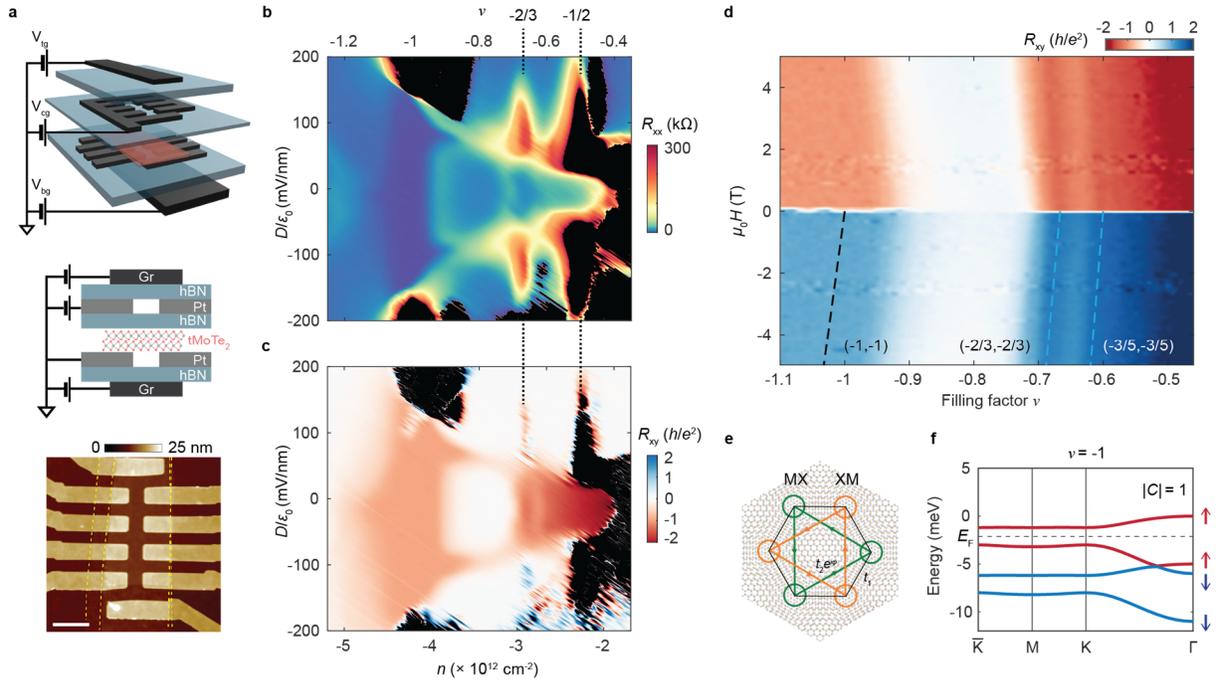

**Figure 1 | Phase diagram of quantized anomalous Hall states. a**, Top: Device schematic with contact gate design. Graphite (Gr) top and bottom gates form a dual gate geometry, while patterned local gates are used to improve the electrical conductivity between platinum contacts and the twisted MoTe$_2$ bilayer. Middle: Side view of the device geometry. Bottom: atomic force microscopy image of device D(3.7°) during fabrication. Dashed lines indicate the edges of the MoTe$_2$ flakes. Scale bar: 2 µm. **b**, Longitudinal ($R_{xx}$) and **c**, Hall ($R_{xy}$) resistance as a function of electric field ($D/\varepsilon_0$) and carrier density ($n$) at 100 mK. The filling factor ($v$) is shown on the top axis. $R_{xx}$, $R_{xy}$ are obtained via the contact scheme detailed in Extended Data Fig. 1 from device D(3.7°). The resistance in black shaded regions cannot be reliably probed due to their highly insulating nature. $R_{xx}$ and $R_{xy}$ are symmetrized and antisymmetrized at $|\mu_0 H|$ = 200 mT, respectively. **d**, Landau fan of unsymmetrized $R_{xy}$ at 100 mK. Black line corresponds to a $C$ = -1 QAH state at $v$ = -1, while the blue lines correspond to the $C$ = -2/3 FQAH state at $v$ = -2/3, and $C$ = -3/5 FQAH state at $v$ = -3/5. **e**, The honeycomb superlattice formed by the two degenerate moiré orbitals localized at the MX (green) and XM (orange) sites. Nearest-neighbor hopping $t_1$ and next-nearest-neighbor complex hopping $t_2 e^{i\varphi}$ between moiré orbitals realize the Kane-Mele model. **f**, Hartree-Fock calculation of topological bands at $v$ = -1 and $D/\varepsilon_0$ = 0. The topmost valence band has a Chern number of $|C|$ = 1 (see Methods). Red and blue curves represent spin up and down electronic bands, respectively. When the Fermi level (grey dashed line) is between the two top valence bands, a QAH insulator is expected. Fractional filling of the topmost valence band results in FQAH states.

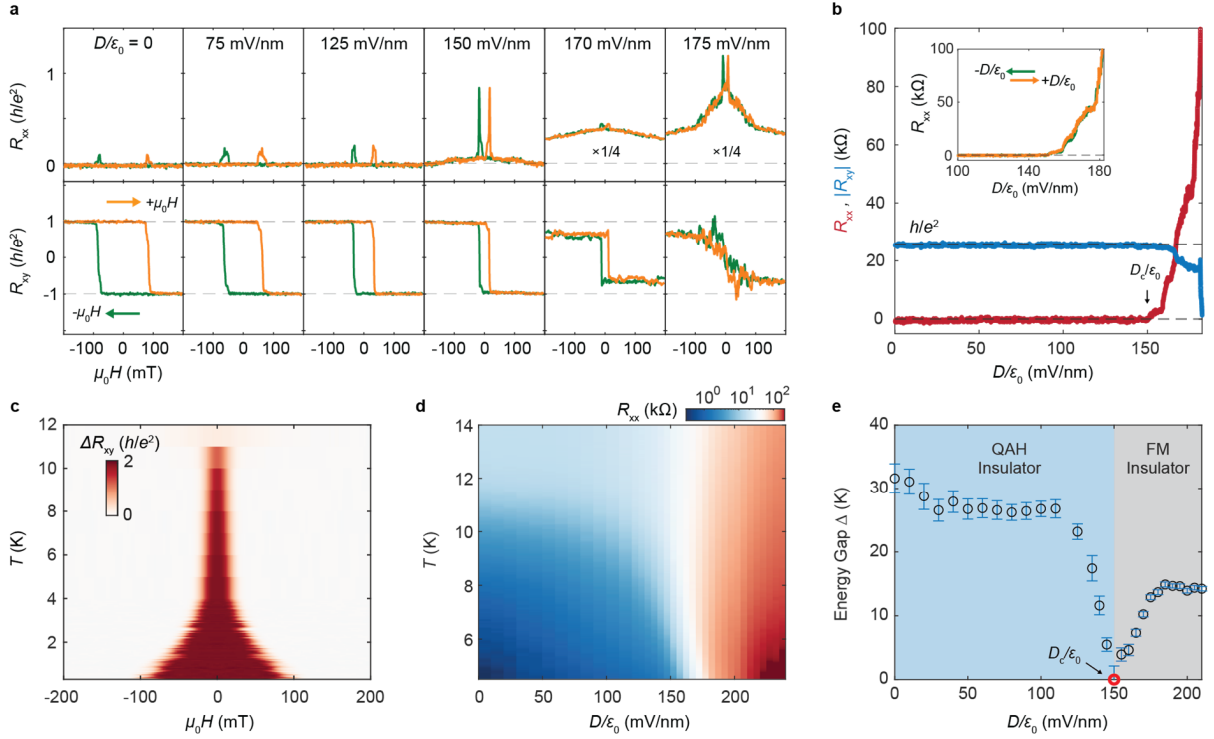

**Figure 2 | Electrically tunable integer QAH state at $v = -1$.** Data are taken from device D(3.9°). **a**, Magnetic field dependent $R_{xx}$ and $R_{xy}$ at selected electric fields $D/\varepsilon_0$ at $T = 100$ mK. Quantized $R_{xy}$ and vanishing $R_{xx}$ at zero magnetic field, as can be observed for $D/\varepsilon_0 = 0$, demonstrate the QAH effect. **b**, $|R_{xy}|$ (blue) and $R_{xx}$ (red) as a function of $D/\varepsilon_0$. $R_{xx}$ and $R_{xy}$ are symmetrized and antisymmetrized at $|\mu_0 H| = 200$ mT, respectively. Inset: $R_{xx}$ versus electric field $D/\varepsilon_0$ swept up and down near the phase transition. **c**, $\Delta R_{xy}$, the hysteretic component of $R_{xy}$, as a function of temperature and magnetic field. $\Delta R_{xy}$ is remains nearly quantized at $2h/e^2$ up to 8 K, demonstrating the robustness of the QAH state. **d**, Intensity plot of $R_{xx}$ versus temperature and electric field. A small magnetic field (100 mT) is applied to suppress magnetic domain fluctuations. The system transitions from a QAH to a trivial insulating phase as $D/\varepsilon_0$ is increased above approximately 150 mV/nm. **e**, Energy gap as a function of electric field extracted from the data in (**d**). Error bars are obtained from fitting variance. The closing and reopening of the gap versus $D/\varepsilon_0$ is evidence for a continuous topological quantum phase transition between the QAH state and topologically trivial correlated insulating state.

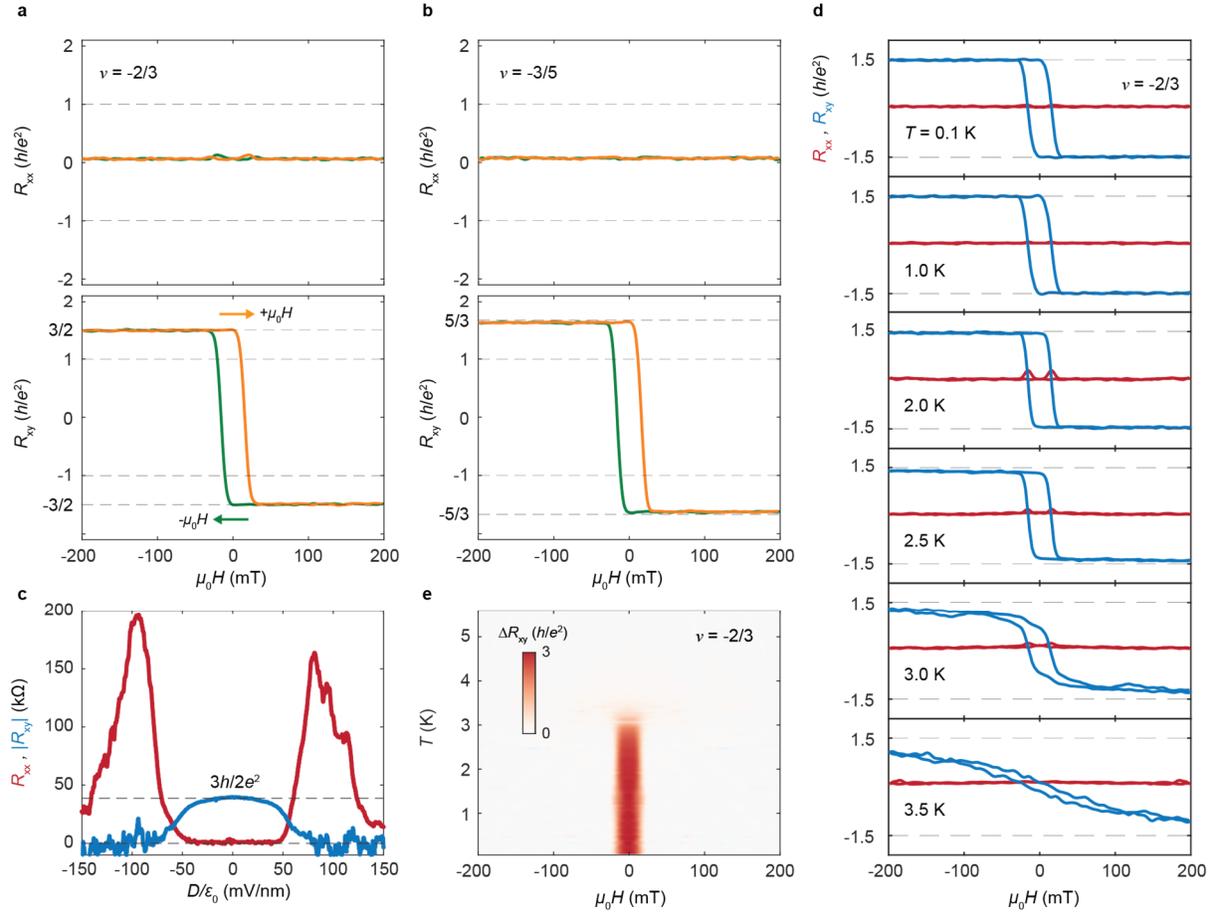

**Figure 3 | Fractionally quantized anomalous Hall effect.** Data are taken from device D(3.7°). **a**, **b**, $R_{xx}$ and $R_{xy}$ versus magnetic field $\mu_0 H$ at **a**, $v = -2/3$ and **b**, $v = -3/5$, with $T = 500$ mK and $D/\varepsilon_0 = 0$. The data in panels (**c-e**) are taken at the $v = -2/3$ FQAH state. **c**, Antisymmetrized $|R_{xy}|$ (blue) and symmetrized $R_{xx}$ (red) at $|\mu_0 H| = 50$ mT as a function of $D/\varepsilon_0$ at $T = 100$ mK. **d**, Magnetic field dependence of $R_{xx}$ and $R_{xy}$ at selected temperatures with an electric field of $D/\varepsilon_0 = 0$. $R_{xy}$ remains nearly quantized at 1 K. **e,** $\Delta R_{xy}$, the hysteretic component of $R_{xy}$, as a function of temperature and magnetic field. $R_{xx}$ and $R_{xy}$ are obtained using the contact scheme detailed in Extended Data Fig. 4.

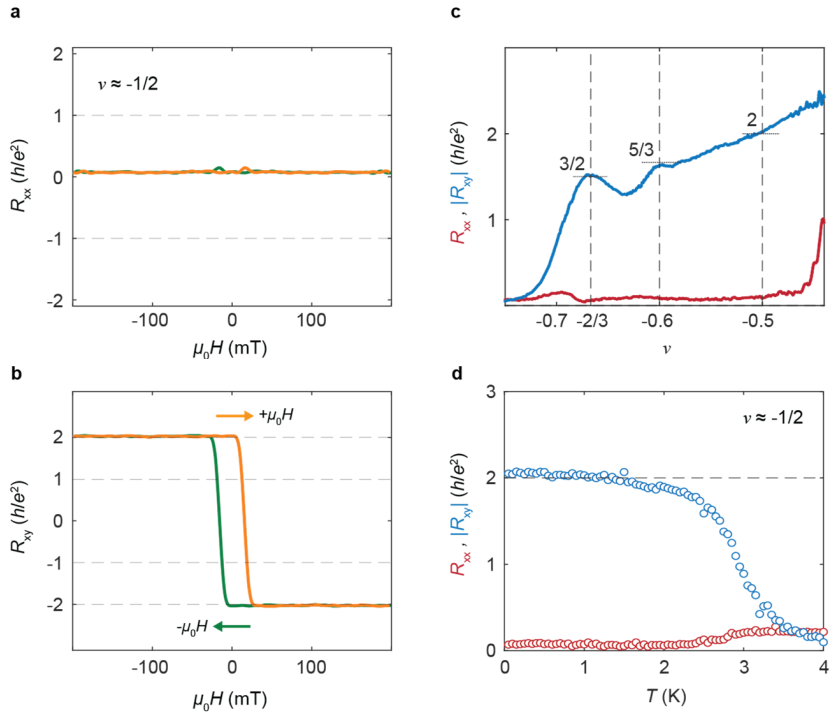

**Figure 4 | Quantum anomalous Hall effect at half filling.** All data are taken at $D/\varepsilon_0 = 0$ and from device D(3.7°). **a**, $R_{xx}$ and **b**, $R_{xy}$ versus magnetic field sweeps at $v \approx -1/2$ and $T = 500$ mK. **c**, Symmetrized $R_{xx}$ (red) and antisymmetrized $R_{xy}$ (blue) at $\pm 50$ mT versus filling factor $v$, at $T \approx 100$ mK. **d**, Temperature dependence of symmetrized $R_{xx}$ and antisymmetrized $R_{xy}$ at $\pm 50$ mT for $v$ near -1/2. A phase transition is evident between 2 and 4 K. $R_{xx}$ and $R_{xy}$ are obtained using the contact scheme detailed in Extended Data Fig. 4.

# Extended Data Figures for

## Observation of Fractionally Quantized Anomalous Hall Effect

**Authors:** Heonjoon Park[1†], Jiaqi Cai[1†], Eric Anderson[1†], Yinong Zhang[1], Jiayi Zhu[1], Xiaoyu Liu[2], Chong Wang[2], William Holtzmann[1], Chaowei Hu[1], Zhaoyu Liu[1], Takashi Taniguchi[5], Kenji Watanabe[6], Jiun-haw Chu[1], Ting Cao[2], Liang Fu[7], Wang Yao[3,4], Cui-Zu Chang[8], David Cobden[1], Di Xiao[2,1], Xiaodong Xu[1,2*]

[1]Department of Physics, University of Washington, Seattle, Washington 98195, USA
[2]Department of Materials Science and Engineering, University of Washington, Seattle, Washington 98195, USA
[3]Department of Physics, University of Hong Kong, Hong Kong, China
[4]HKU-UCAS Joint Institute of Theoretical and Computational Physics at Hong Kong, China
[5]Research Center for Materials Nanoarchitectonics, National Institute for Materials Science, 1-1 Namiki, Tsukuba 305-0044, Japan
[6]Research Center for Electronic and Optical Materials, National Institute for Materials Science, 1-1 Namiki, Tsukuba 305-0044, Japan
[7]Department of Physics, Massachusetts Institute of Technology, Cambridge, Massachusetts 02139, USA
[8]Department of Physics, The Pennsylvania State University, University Park, PA 16802, USA
[†] These authors contributed equally to the work.
*Corresponding author's email: xuxd@uw.edu

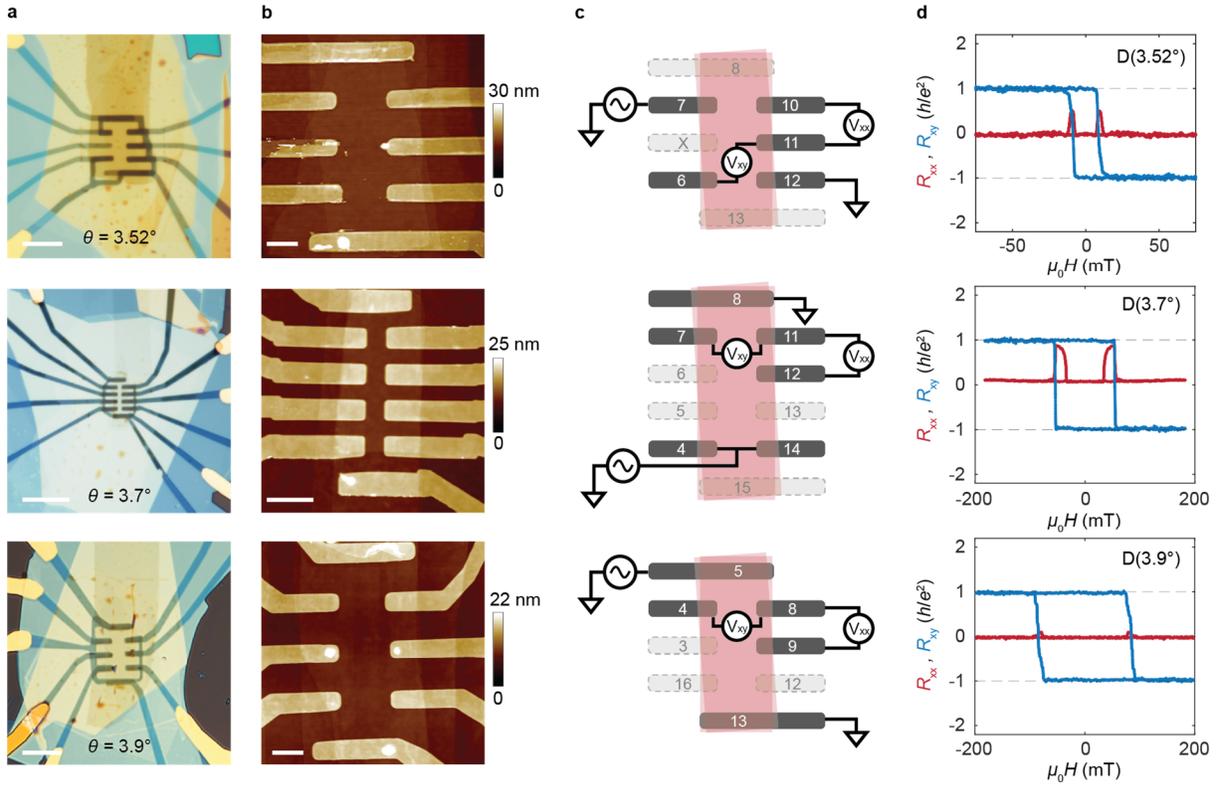

**Extended Data Figure 1 | Reproducibility of the quantum anomalous Hall effect. a**, Optical microscope images of three Hall bar devices studied in this work. Twist angles are indicated. Scale bar: 10 μm. **b**, Atomic force microscopy image of the device before contact gate fabrication. Scale bar: 2 μm. **c**, Contact geometry used in the main figures for the respective devices unless otherwise specified. The electrodes in dashed lines are floated. **d**, Magnetic field dependence of $R_{xx}$ and $R_{xy}$ at 100 mK. The quantization of $R_{xy}$ and concomitant nearly vanishing $R_{xx}$ demonstrates the robustness of the quantum anomalous Hall effect in tMoTe$_2$ across a large range of twist angles.

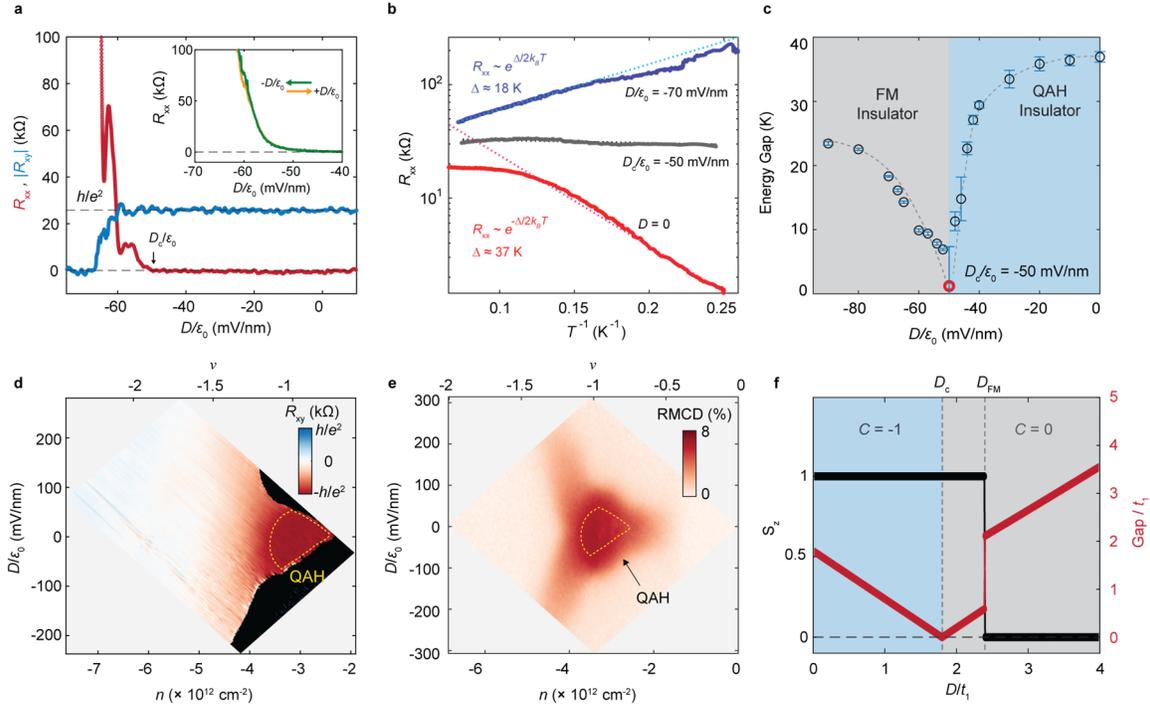

**Extended Data Figure 2 | Electric field dependence of $v = -1$ QAHE in device D(3.52°).** All data in the main panel are taken at 500 mK. **a**, Antisymmetrized $|R_{xy}|$ (blue) and symmetrized $R_{xx}$ (red) at $|\mu_0 H| = 100$ mT as a function of electric field ($D/\varepsilon_0$) at $v = -1$. As $D/\varepsilon_0$ is decreased below $D_c/\varepsilon_0 \approx -50$ mV/nm, $R_{xy}$ quickly drops from the quantized value while $R_{xx}$ increases rapidly to above 100 kΩ. This demonstrates an electric field induced transition from a QAH insulator to topologically trivial correlated insulator. Inset: Symmetrized $R_{xx}$ at $|\mu_0 H| = 100$mT versus electric field $D$ swept down and up near the phase transition. The absence of hysteresis implies a continuous topological phase transition. The data in the inset is taken at 1.6 K to minimize electrical noise from the contacts. **b**, Illustration of fitting used to extract the energy gap Δ at selected electric field values. The Arrhenius equation and extracted Δ are shown for $D/\varepsilon_0 = -70$ and 0 mV/nm. At the critical field $D_c/\varepsilon_0 \approx -50$ mV/nm, the longitudinal resistance is nearly constant vs temperature. **c**, Energy gap as a function of electric field. Error bars are obtained from fitting variance. Grey dashed lines are guides to the eye. The closing and reopening of the gap versus $D/\varepsilon_0$ are evidence for a continuous topological quantum phase transition between the QAH state and a topologically trivial correlated insulator. **d**, Antisymmetrized $R_{xy}$ at magnetic field $|\mu_0 H| = 100$ mT as a function $D/\varepsilon_0$ and carrier density ($n$) at 1.6 K. The filling factor ($v$) is shown on the top axis. Black regions denote areas with resistance too large to be reliably measured. The yellow dashed line bounds the region where $R_{xy}$ is > 95 % of $h/e^2$. **e**, Reflective magnetic circular dichroism (RMCD) signal versus $D/\varepsilon_0$ and $v$ at $\mu_0 H = 100$ mT and 1.6 K. The comparison with panel (**d**) shows that the critical electric field $D_{FM}/\varepsilon_0$ for suppressing the ferromagnetic state is larger than $D_c/\varepsilon_0$ for the QAH insulator (the yellow dashed line is as in panel **d**). Therefore, the topologically trivial state is a ferromagnetic insulator. **f**, Hartree-Fock calculations of the out-of-plane spin ($S_z$) and energy gap normalized to the hopping $t_1$ as a function of electric field. The system is in a topologically trivial ferromagnetic insulating state between $D_c$ and $D_{FM}$, and becomes non-valley polarized and hence non-ferromagnetic above $D_{FM}$. The calculated Chern number is $C = -1$(blue region) and 0 (grey region) for below and above $D_c$, respectively.

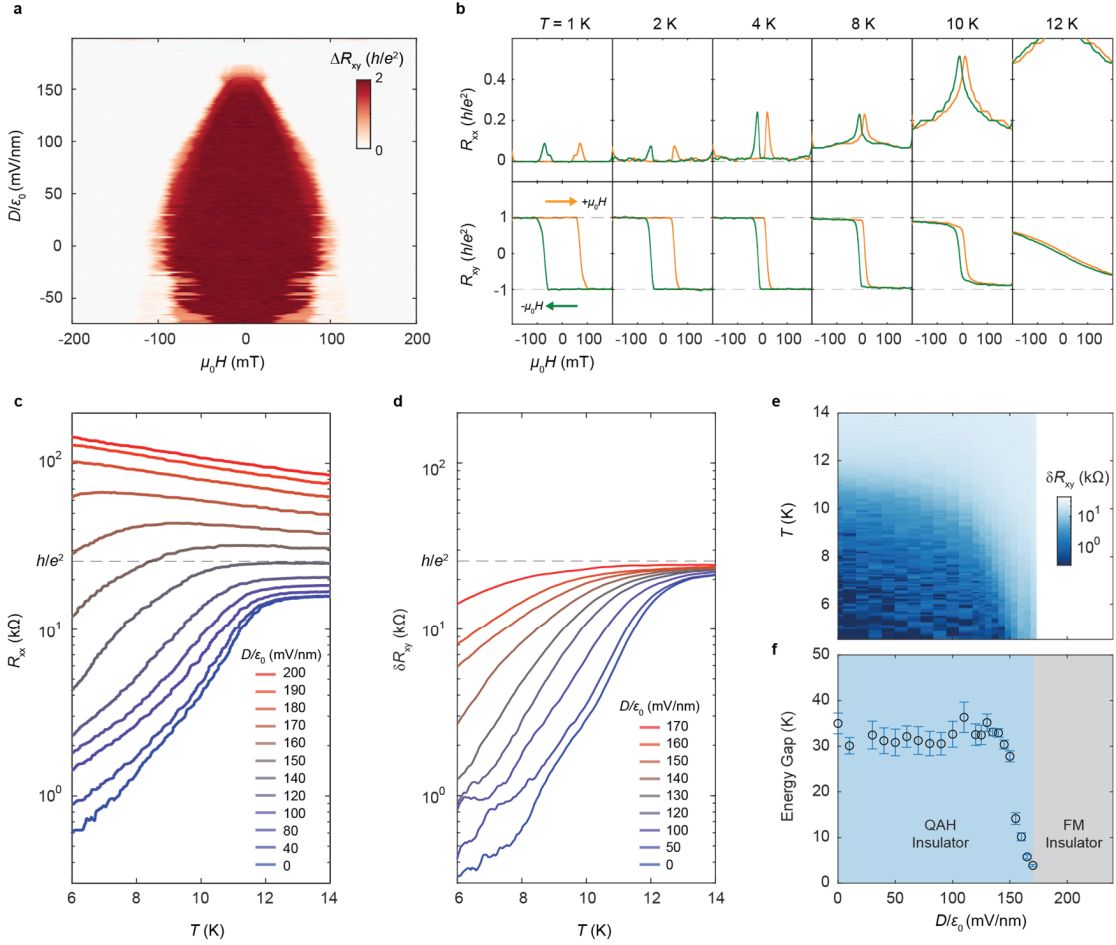

**Extended Data Figure 3 | Electric field and temperature dependent QAH effect at *v* = -1.** Data are taken from device D(3.9°). **a,** $\Delta R_{xy}$ versus magnetic field and electric field. Here, $\Delta R_{xy}$ is obtained by taking the difference of $R_{xy}$ between sweeping magnetic field up and down. $\Delta R_{xy}$ starts to vanish around 150 mV/nm, signifying a topological phase transition. Large negative electric field values cannot be reached due to gate limits. **b,** Magnetic field dependence of $R_{xx}$ and $R_{xy}$ at selected temperatures. $R_{xy}$ is nearly quantized even at 8 K. **c,** $R_{xx}$ as a function of temperature at selected electric fields. $R_{xx}$ is unsymmetrized and taken at 100 mT to avoid magnetic instability. **d,** Temperature dependent $\delta R_{xy} = |h/e^2 - R_{xy}|$ at selected electric fields, where $R_{xy}$ is the antisymmetrized Hall resistance at ±100 mT. **e,** Color plot of $\delta R_{xy}$ showing a phase transition near 150 mV/nm. **f,** Estimated thermal activation gap from the data in (**e**). An energy gap of 35(2) K near zero electric field can be extracted, which is consistent with the value obtained from $R_{xx}$ measurements, 32(2) K (see main text Figure 2e).

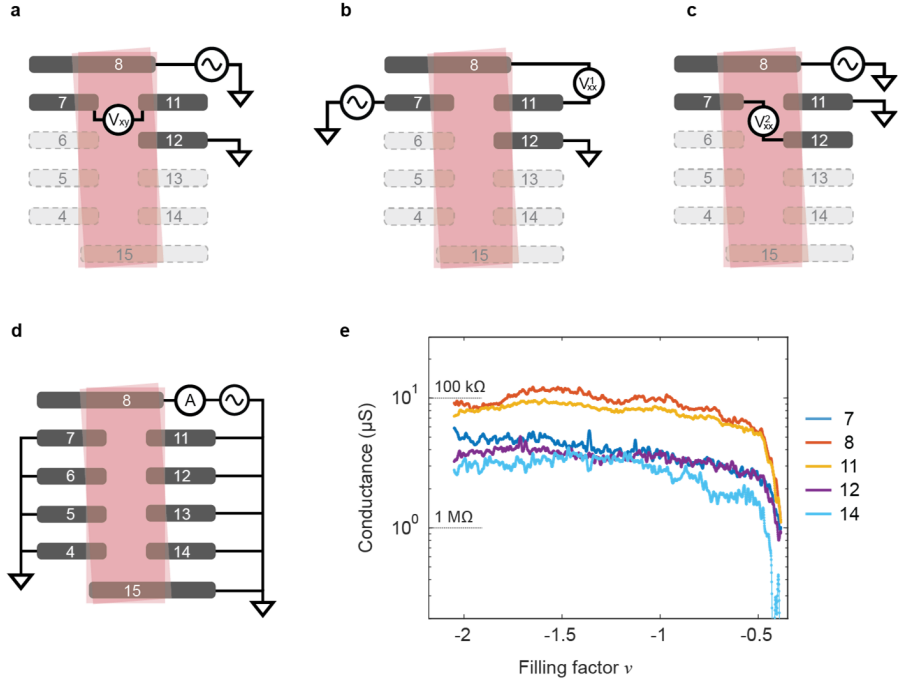

**Extended Data Figure 4 | Measurement configurations for Figs. 3 and 4 in the main text, and characterization of electrical contacts for device D(3.7°).** We find that the contacts 7, 8, 11, and 12 isolate a single FQAH domain. **a,** Configuration for measuring $R_{xy}$. Current flows from contacts 8 to 12, and Hall voltage is measured between contacts 7 and 11. **b** and **c**, are reciprocal configurations for measuring $R_{xx}$. The symmetrized data $(R^1_{xx}+R^2_{xx})/2$ from these configurations result in $R_{xx}$. The electrodes in dashed lines are floated. **d,** Configuration for two terminal measurements sourcing current from one contact while keeping all remaining contacts grounded (using contact 8 as a representative case). **e,** Doping dependence of contact conductance at 500 mK for the contacts used in the experiment. All contacts remain under 500 kΩ until resistance starts to increase near $v = -1/2$ as the system enters an insulating state.

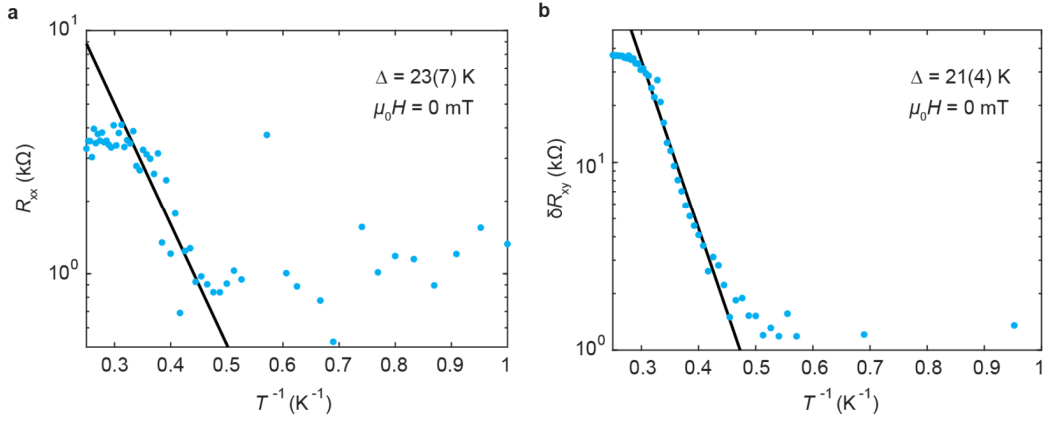

**Extended Data Figure 5 | Estimation of energy gap of the -2/3 FQAH state.** Data are taken from device D(3.7°) using the contact configuration in Extended Data Fig. 4. **a,** Arrhenius plot of $R_{xx}$. Solid line is the fit to equation $R_0 e^{-\Delta/2k_B T}$. **b**, Arrhenius plot of $\delta R_{xy} = (3h/2e^2 - R_{xy})$. Solid line is the fit to $R_0 e^{-\Delta/k_B T}$. Extracted energy gaps from $R_{xx}$ and $R_{xy}$ measurements are similar, with values of 23(7)K and 21(4)K, respectively.

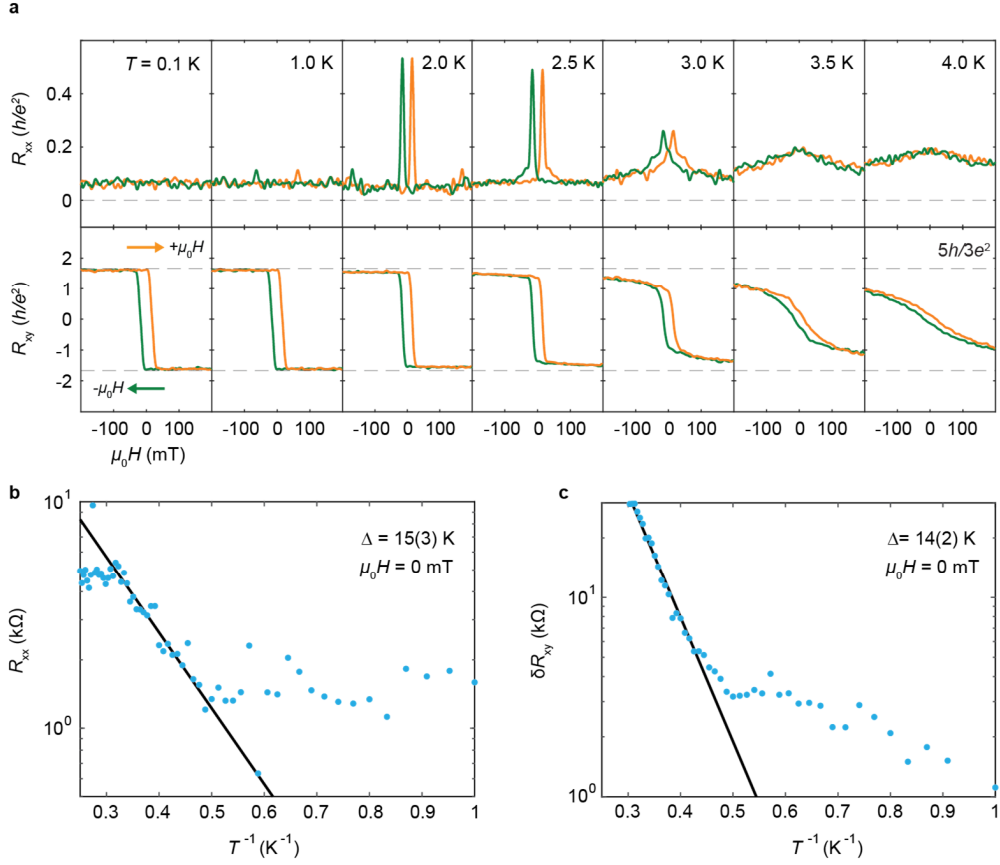

**Extended Data Figure 6 | Temperature dependent measurements of the -3/5 state.** Data are taken from device D(3.7°). **a,** Magnetic field dependence of $R_{xx}$ (top) and $R_{xy}$ (bottom) at selected temperatures and $D/\varepsilon_0 = 0$. $R_{xy}$ is measured in the configuration described in Extended data Fig. 4a, while $R_{xx}$ is obtained from the configurations in Extended Data Figs. 4b and c. **b,** Arrhenius plot of $R_{xx}$. Solid line is the fit to equation $R_0 e^{-\Delta/2k_B T}$. **c,** Arrhenius plot of $\delta R_{xy} = (5h/3e^2 - R_{xy})$. Solid line is the fit to $R_0 e^{-\Delta/k_B T}$. Extracted energy gaps from $R_{xx}$ and $R_{xy}$ measurements are similar, with values of 15(3) K and 14(2) K, respectively.

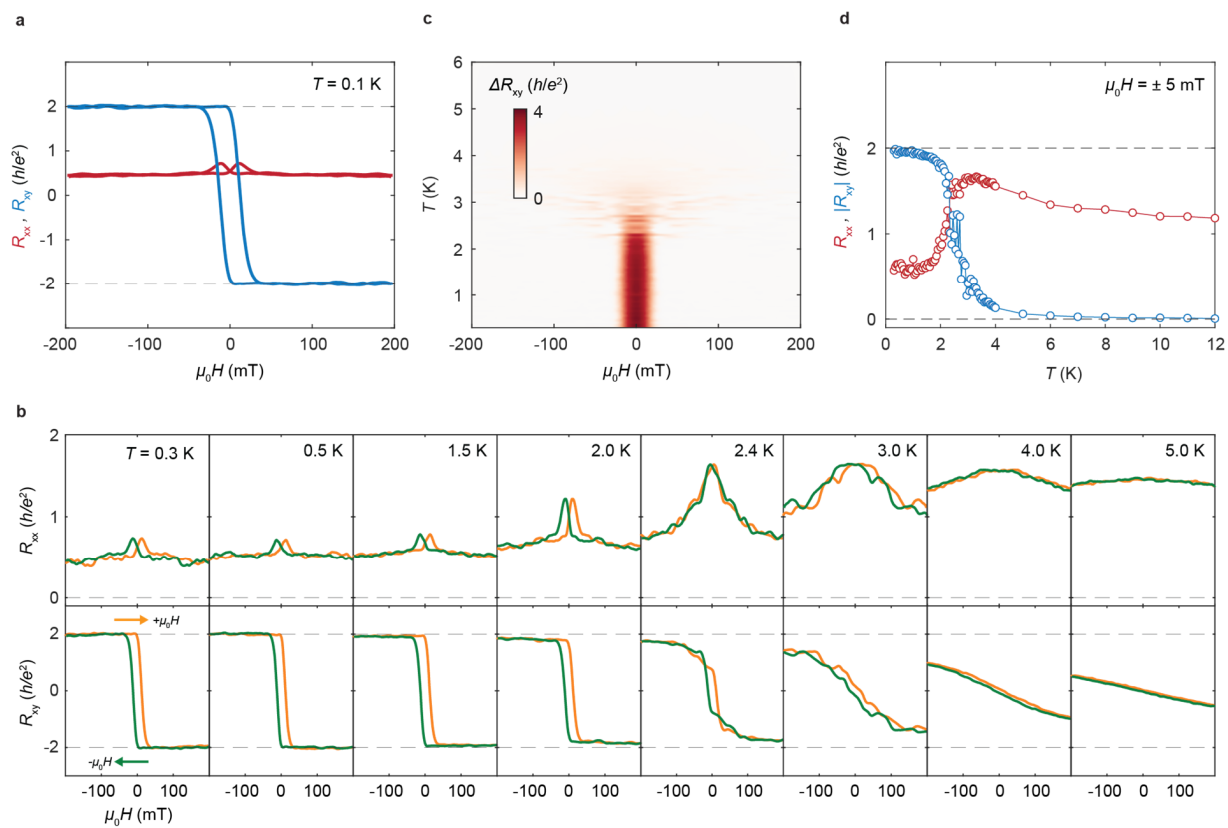

**Extended Data Figure 7 | -1/2 QAH state in an additional device D(3.9⁰). a,** $R_{xx}$ and $R_{xy}$ versus magnetic field near $v = -1/2$, $D/\varepsilon_0 = 0$, and at a temperature of 100 mK. **b,** Magnetic field dependent $R_{xx}$ (top) and $R_{xy}$ (bottom) at selected temperatures. **c,** $\Delta R_{xy}$ versus magnetic field and temperature. Here, $\Delta R_{xy}$ is the difference between $R_{xy}$ for magnetic field swept up and down. **d,** Temperature dependent symmetrized $R_{xx}$ and antisymmetrized $R_{xy}$ at ±5 mT and $v$ near -1/2. A phase transition between 2-4 K is evident.

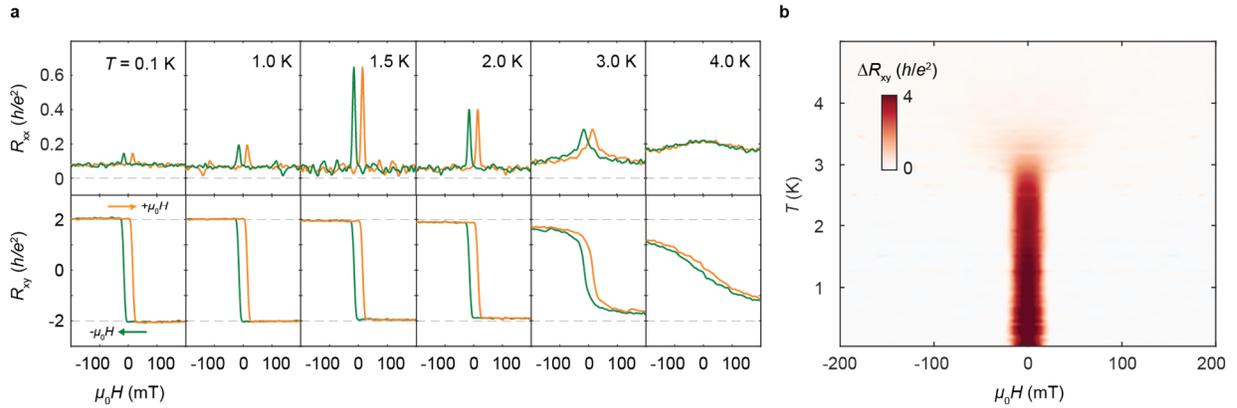

**Extended Data Figure 8 | Temperature dependent transport for -1/2 QAH state from D(3.7⁰).** Data are taken from device D(3.7°) using the contact configuration in Extended Data Fig. 4. **a**, Magnetic field dependent $R_{xx}$ and $R_{xy}$ at different temperatures near $v = -1/2$ and $D/\varepsilon_0 = 0$. **b**, $\Delta R_{xy}$ versus magnetic field and temperature. Here, $\Delta R_{xy}$ is the difference between $R_{xy}$ for magnetic field swept up and down.

9